\documentclass{aa}

\usepackage{graphicx}
\usepackage{times}

\begin{document}
\sloppy

\thesaurus{06(08.09.1; 08.09.2 \object{AFGL 2290}; 08.16.4; 08.13.2;
           08.03.4; 13.09.6)} 

\title{High--resolution speckle masking interferometry and radiative
       transfer modeling of the oxygen--rich AGB star
       \object{AFGL 2290}\thanks{ 
         Based on data collected at the 6~m telescope of the Special
         Astrophysical Observatory in Russia}} 

\author{A.~Gauger\inst{1} \and Y.Y.~Balega\inst{2} \and
        P.~Irrgang\inst{1} \and R.~Osterbart\inst{1} \and
        G.~Weigelt\inst{1}} 

\institute{Max--Planck--Institut f{\"u}r Radioastronomie,
           Auf dem H{\"u}gel 69, D--53121 Bonn, Germany
          \and
           Special Astrophysical Observatory, Nizhnij Arkhyz,
           Karachai-Cherkesia, 357147, Russia}

\offprints{R.~Osterbart\\\mbox{(osterbart@mpifr-bonn.mpg.de)}}

\date{Received 28 September 1998 / Accepted 10 March 1999}

\titlerunning{Speckle masking interferometry and
              radiative transfer modeling of \object{AFGL 2290}} 
\authorrunning{Gauger et al.}

\maketitle

\begin{abstract}
We present the first diffraction--limited speckle masking observations
of the oxygen--rich AGB star \object{AFGL 2290}. The speckle
interferograms were recorded with the Russian 6~m SAO telescope. At
the wavelength $2.11\,{\rm\mu m}$ a resolution of 75 milli--arcsec
(mas) was obtained. The reconstructed diffraction--limited image
reveals that the circumstellar dust shell (CDS) of \object{AFGL 2290}
is at least slightly non--spherical. The visibility function shows
that the stellar 
contribution to the total $2.11\,{\rm\mu m}$ flux is less than $\sim
40 \%$, indicating a rather large optical depth of the circumstellar
dust shell. The 2--dimensional Gaussian visibility fit yields a
diameter of \object{AFGL 2290} at $2.11\,{\rm\mu m}$ of
43~mas$\times$51~mas, which corresponds to a diameter of
42~AU$\times$50~AU for an adopted distance of 0.98~kpc.

Our new observational results provide additional constraints on the
CDS of \object{AFGL 2290}, which supplement the information from the
spectral energy distribution (SED). To determine the structure and the
properties of the CDS we have performed radiative transfer
calculations for spherically symmetric dust shell models.  
The observed SED approximately at phase 0.2 can be well reproduced at
all wavelengths by a model with $T_{\rm eff}=2000\,{\rm K}$, a dust
temperature of 800~K at the inner boundary $r_{1}$, an optical depth
$\tau_{V}=100$ and a radius for the single--sized grains of $a_{\rm
gr}=0.1\,{\rm\mu m}$. However, the $2.11\,{\rm\mu m}$ visibility of
the model does not match the observation.

Exploring the parameter space, we found that grain size is the key
parameter in achieving a fit of the observed visibility while retaining
the match of the SED, at least partially. Both the slope and the
curvature of the visibility strongly constrain the possible grain
radii. On the other hand, the SED at longer wavelengths, the silicate
feature in particular, determines the dust mass loss rate and, thereby,
restricts the possible optical depths of the model.
With a larger grain size of $0.16\,{\rm\mu m}$ and a higher
$\tau_{V}=150$, the observed visibility can be reproduced preserving
the match of the SED at longer wavelengths. Nevertheless, the model
shows a deficiency of flux at short wavelengths, which is attributed
to the model assumption of a spherically symmetric dust distribution,
whereas the actual structure of the CDS around \object{AFGL 2290} is
in fact non--spherical. 
Our study demonstrates the possible limitations of dust shell models
which are constrained solely by the spectral energy distribution, and
emphasizes the importance of high spatial resolution observations for
the determination of the structure and the properties of circumstellar
dust shells around evolved stars.


\end{abstract}


\section{Introduction}

\object{AFGL 2290} (OH~39.7+1.5, IRAS~18560+0638, V1366 Aql) belongs
to the group of type II OH/IR stars, which can be defined as infrared
point sources with a maximum of the spectral energy distribution (SED)
around $6 - 10\,{\rm\mu m}$, with the $9.7\,{\rm\mu m}$ silicate
band in absorption, and with OH maser emission in the 1612 MHz line
(Habing \cite{Hab96}). Most of these objects show a long-period
variability in the infrared and the OH maser emission (Engels
\cite{Eng82}; Herman \& Habing \cite{HerHab85}), although also a small
fraction either varies irregularly with small amplitude or does not
vary at all. OH/IR stars are surrounded by massive circumstellar
envelopes composed of gas and small solid particles (dust, grains).
These circumstellar dust shells (CDS) are produced by the ejection of
matter at large rates ($\dot{M}>10^{-7}$ ${\rm M_{\odot}\,yr^{-1}}$)
and low velocities ($\sim 15\,{\rm kms^{-1}}$), and in some cases they
totally obscur the underlying star. Based on the luminosities ($\sim
10^{4}\,\rm L_{\odot}$), the periods (500d to 3000d) and bolometric
amplitudes ($\sim 1$~mag), the kinematical properties and galactic
distribution, the majority of OH/IR stars are highly evolved low-- and
intermediate--mass stars populating the asymptotic giant branch (AGB)
(Habing \cite{Hab96}). They extend the sequence of optical Mira
variables to longer periods, larger optical depths and higher mass
loss rates (Engels et al.\ \cite{EKSS83}; Habing \cite{Hab90}; Lepine
et al.\ \cite{LOE95}). 

The improvements of the observational techniques, especially at
infrared wavelengths, and the elaboration of increasingly
sophisticated theoretical models have provided a wealth of new
information on the structure, the dynamics, and the evolution of the
atmospheres and circumstellar shells of AGB stars, although many
details still remain to be clarified (see the review by Habing
\cite{Hab96}). A general picture has become widely accepted in which
both the large amplitude pulsations and the acceleration by radiation
pressure on dust contribute to the mass loss phenomenon for AGB stars. 
From observations, correlations are found between the period and the
infrared excess (indicating the mass loss rate)
(DeGioia--Eastwood et al.\ \cite{DHGG81}; Jura \cite{Jur86}), and
between the period and the terminal outflow velocity (Heske
\cite{Hes90}). On the theoretical side, hydrodynamical models showed
that due to the passage of shocks generated by the stellar pulsation
the atmosphere is highly extended, thus enabling dust formation and
the subsequent acceleration of the matter (Wood \cite{Wood79}; Bowen
\cite{Bow88}). The inclusion of a detailed treatment of dust formation
revealed a complex interaction between pulsation and dust formation,
which results e.g.\ in a layered dust distribution and affects the
derived optical appearance (Fleischer et al.\ \cite{fgs92},
\cite{fgs95}; Winters et al.\ \cite{wfgs94}, \cite{wfgs95}). 

Until now most interpretations of observations as well as most
theoretical models are based on the assumption of a spherically
symmetric dust shell, often motivated by the circularity of the OH
maser maps. However, observations show that some objects have
substantial deviations from spherical symmetry (e.g.\ Dyck et al.\
\cite{DZLB84}; Kastner \& Weintraub \cite{KaWe94}; Weigelt  et al.\
\cite{WBBFOW98}). This suggests that the asymmetries observed in many
post--AGB objects and planetary nebulae (cf.\ Iben \cite{Ib95}) may
already start to develop during the preceding AGB phase, which
provides new challenges for the modeling of the mechanisms and processes
determining the structure of the dust shells around AGB stars.

High spatial resolution observations can yield direct information on
important properties of the dust shells around AGB stars, such as the
dimensions and geometry of the shell. Therefore, such observations
contribute additional strong constraints for the modeling of these
circumstellar environments, which supplement the information from the
spectral energy distribution.
Measurements of the visibility at near-IR wavelengths, for example,
can be used to determine the radius of the onset of dust formation as
well as to constrain the dominant grain size (Groenewegen
\cite{Groe97}). To gain information on details of the spatial
structure, in particular on asymmetries and inhomogeneties of the CDS,
the interferometric imaging with large single--dish telescopes is
especially well suited because one observation provides all spatial
frequencies up to the diffraction limit of the telescope and for all
position angles simultaneously, allowing the reconstruction of true
images of the object.

We have chosen \object{AFGL 2290} for our study because it represents
a typical obscured OH/IR star with a high mass loss rate, whose
location is not too far away from us. The distance to
\object{AFGL 2290} can be determined directly with the phase lag
method (cf.\ Jewell et al.\ \cite{JEWS79}), which gives $D =
0.98\,{\rm kpc}$ (van Langevelde et al.\ \cite{LHS90}). For the
bolometric flux at earth a value of $f_{\rm b} \sim
2.4\,10^{-10}\,{\rm W m^{-2}}$ is derived by van der Veen \& Rugers
(\cite{VeRu89}) from infrared photometry between $1\,{\rm\mu m}$ and
$12\,{\rm\mu m}$ and the IRAS fluxes. At $0.98\,{\rm kpc}$ the
luminosity is $L=7200\,{\rm L_{\odot}}$, which is within the typical
range for an oxygen--rich AGB star. The long period of $P=1424\,{\rm
d}$ determined from the variation of the OH maser (Herman \& Habing
\cite{HerHab85}) and the high mass loss rate suggest, that the star is
in a late phase of its AGB evolution. 

So far, Chapman \& Wolstencroft (\cite{ChWo87}) reported the only high
angular--resolution infrared observations of \object{AFGL 2290}. From
1--dimensional slit--scan speckle interferometry with the UKIRT 3.8~m
telescope at $3.8\,{\rm\mu m}$ and $4.8\,{\rm\mu m}$ they derive
1--dimensional visibilities and determine Gaussian FWHM diameters.
Radiative transfer models for the \object{AFGL 2290} dust shell have
been presented by Rowan--Robinson (\cite{RowRob82}), Bedijn
(\cite{Bed87}), Suh (\cite{Suh91}) and recently by Bressan et al.\
(\cite{BGS98}). These models yield dust shell properties within the
typical range of OH/IR stars, e.g.\ a dust mass loss rate of about
$4\,10^{-7}\,{\rm M_{\odot} yr^{-1}}$ (Bedijn \cite{Bed87}; Bressan et
al.\ \cite{BGS98}), or an optical depth at $9.7\,{\rm\mu m}$ of about
10 (Bedijn \cite{Bed87}; Suh \cite{Suh91}). However, none of these 
studies includes constraints from high spatial resolution infrared
measurements.

In Sect.~\ref{observations} we present the results of our speckle
masking observations of \object{AFGL 2290}. The approach for the
radiative transfer modeling is described in
Sect.~\ref{radtrf-approach} comprising a short description of the
code, the selection of the photometric data and a discussion of input
parameters for the models. In Sect.~\ref{radtrf-modeling} we present
the results of the radiative transfer modeling starting with the
discussion of a model, which yields a good fit of the observed SED at
all wavelengths but does not reproduce the observed $2.11\,{\rm\mu m}$
visibility. In search of an improved model the changes of the
resulting SED and visibility under variations of the input parameters
are investigated in the following sections. We finish the paper with a
summary of the results and our conclusions in Sect.~\ref{summary}.


\section{Speckle masking observations}
\label{observations}

The \object{AFGL 2290} speckle data presented here were obtained with
the Russian 6~m telescope at the Special Astrophysical Observatory
(SAO) on June 14 and 16, 1998. We recorded a total number of 1200
speckle interferograms of \object{AFGL 2290} (600 on June 14 and 600
on June 16) and 2400 speckle interferograms of the unresolved
reference star HIP~93260 (1200 on each of the two nights) with our
256$\times$256 pixel NICMOS~3 camera through an interference filter
with center wavelength 2.11\,$\mu$m and a bandwidth of 0.192\,$\mu$m.
The exposure time per frame was 100~ms, the pixel size was 30.61~mas
and the field of view $7\farcs8\times7\farcs8$. The $2.11\,{\rm\mu m}$
seeing was about $\sim 1\farcs2$. 
A diffraction-limited image of \object{AFGL 2290}
was reconstructed from the speckle data by the speckle masking
bispectrum method (Weigelt \cite{Weig77}; Lohmann et al.\
\cite{LohmWeWi83}; Weigelt \cite{Weig91}). The process includes the
calculation of the average power spectrum and of the average
bi--spectrum and the subtraction of the detector noise terms from
those. The modulus of the object Fourier transform was determined with
the speckle interferometry method (Labeyrie \cite{Lab70}). The Fourier
phase was derived from the bias--compensated average bispectrum.

\begin{figure}\unitlength 1cm
\resizebox{\hsize}{!}{\includegraphics{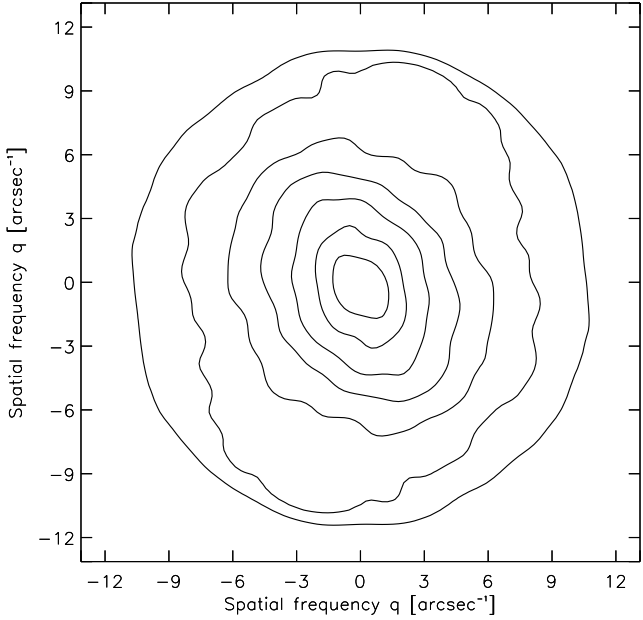}}
\caption{
   Two--dimensional $2.11\,{\rm\mu m}$ visibility function of
   \object{AFGL 2290} derived from the speckle interferograms. 
   The contour levels are plotted from 20\% to 80\% of the peak
   value in steps of 10\%.} 
\label{Powerspectrum}
\end{figure}
 
\begin{figure}\unitlength 1cm
\resizebox{\hsize}{!}{\includegraphics{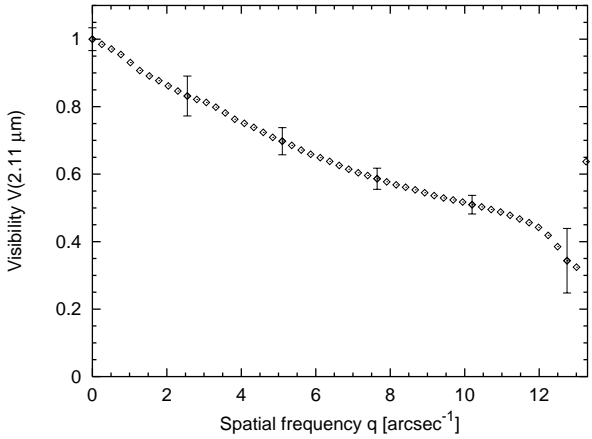}}
\caption{
   Azimuthally averaged $2.11\,{\rm\mu m}$ visibility function of
   \object{AFGL 2290} and errorbars.} 
\label{Visibility}
\end{figure}

Figures \ref{Powerspectrum} and \ref{Visibility} show the visibility
function of \object{AFGL 2290} at $2.11\,{\rm\mu m}$. The azimuthally
averaged visibility decreases steadily to values below $\sim 0.40$ of
the peak visibility at the diffraction cut--off frequency ($13.5\,{\rm
 arcsec^{-1}}$). Thus, the circumstellar dust shell is almost totally
resolved, and 
the contribution of the unresolved stellar component to the 
monochromatic flux at $2.11\,{\rm\mu m}$ must be less than $\sim
40\,\%$, suggesting a rather high optical depth at this wavelength. 
In order to derive diameters for the dust shell, the object visibility
function was fitted with an elliptical Gaussian model visibility
function within a range of $1.5\,{\rm arcsec^{-1}}$ up to $7.5\,{\rm
arcsec^{-1}}$. 
We obtain a Gaussian fit diameter of 43~mas$\times$51~mas
for \object{AFGL 2290} corresponding to 42~AU$\times$50~AU for an 
adopted distance of 0.98~kpc or $5.7\,r_{*}\times6.8\,r_{*}$ for an
adopted distance of 0.98~kpc and an adopted stellar radius of
$r_{*}=7.5\,{\rm mas}$ (cf.\ Sect.~\ref{best-fit-sed}), 
respectively. 

\begin{figure}\unitlength 1cm
\resizebox{\hsize}{!}{\includegraphics{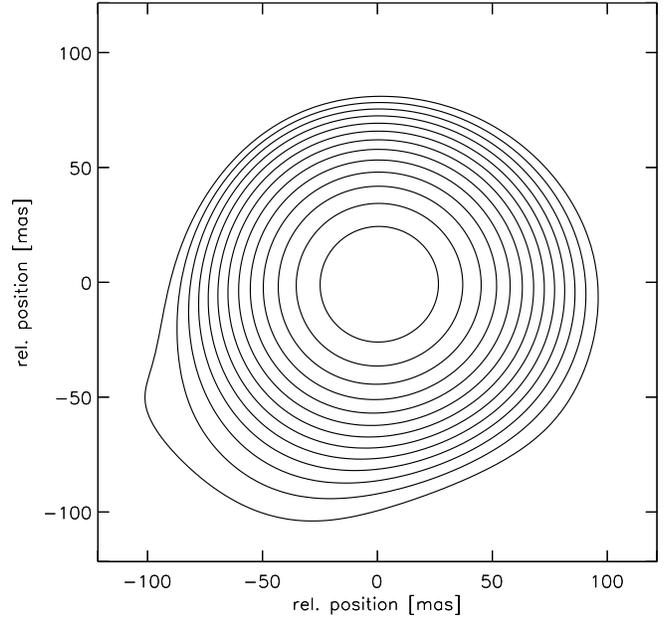}}
\caption{Diffraction--limited $2.11\,{\rm\mu m}$ speckle masking image
   of the \object{AFGL 2290}. North is at the top and east to the left. 
   The contours level intervals are 0.25~mag. The lowest contour level
   is 3.25~mag fainter than the peak intensity.}
\label{object-intensity}
\end{figure}

\begin{figure}\unitlength 1cm
\resizebox{\hsize}{!}{\includegraphics{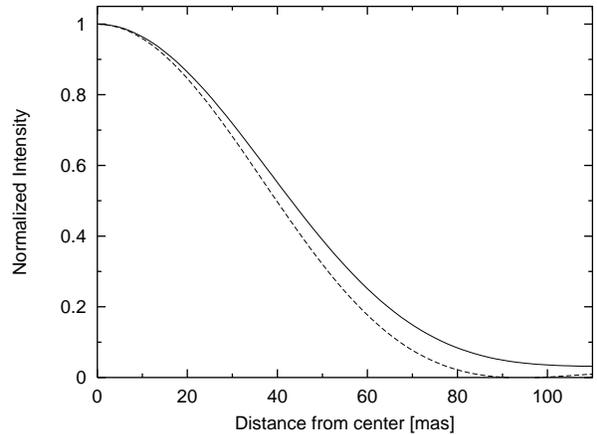}}
\caption{Azimuthally averaged image of \object{AFGL 2290} (solid line)
   and of the unresolved reference star HIP~93260 (dashed line).}
\label{Intensitycuts}
\end{figure}

Figure \ref{object-intensity} shows the reconstructed $2.11\,{\rm\mu
 m}$ speckle masking image of \object{AFGL 2290}. The resolution is
75~mas. Figure \ref{Intensitycuts} shows the azimuthally averaged
images derived from the reconstructed 2--dimensional images of
\object{AFGL 2290} and the reference star HIP~93260. In the
2-dimensional \object{AFGL 2290} image a deviation from
spherical symmetry can be recognized. The intensity contours are
elongated in the south--eastern direction along an axis with a position
angle of $130^{\circ}$.


\section{The radiative transfer modeling approach}
\label{radtrf-approach}

\subsection{The radiative transfer code}
\label{radtrf-code}

The radiative transfer calculations are performed with the code DUSTY
developed by Ivezi{\'c} et al.\ (\cite{INE97}), which is publicly
available. 
The program solves the radiative transfer problem for a spherically
symmetric dust distribution around a central source of radiation and
takes full advantage of the scaling properties inherent in the
formulation of the problem. The formulation of the radiative transfer
problem, the model assumptions and the scaling properties are
described in detail by Ivezi{\'c} \& Elitzur (\cite{IE97}). Therefore,
we give only a brief discussion here.
The problem under consideration is a spherically symmetric dust
envelope with a dust free inner cavity surrounding a central source
of radiation. This geometry is not restricted to the dust shell of a
single star. It can as well describe a dust envelope around a group
of stars (e.g\ a binary) or even around a galactic nucleus. The radial
dependence of the dust density between the inner and outer boundary
can be chosen arbitrarily. 
To arrive at a scale invariant formulation two assumptions are
introduced: i) the grains are in radiative equilibrium with the
radiation field, and ii) the location of the inner boundary $r_1$ of
the dust envelope is controlled by a fixed temperature $T_1$ of the
grains at $r_1$. Due to radiative equilibrium this temperature is
determined by the energy flux at $r_1$, which in turn is controlled
by the energy flux from the central source via the radiative transfer 
through the dusty envelope. Then prescribing the dust temperature at
$r_1$ is equivalent to specifying the bolometric flux at the inner
boundary, and the only relevant property of the input radiation is its
spectral shape (Ivezi{\'c} \& Elitzur \cite{IE97}). Similarly, if the
overall optical depth of the dust envelope at some reference
wavelength is prescribed, only dimensionless, normalized distributions
describing the spatial variation of the dust density and the
wavelength dependence of the grain optical properties enter into the
problem.  

This formulation of the radiative transfer problem for a dusty
envelope is well suited for model fits of IR observations, because it
minimizes the number of independent model parameters. The input
consists of: 
\begin{itemize}
\item the spectral shape of the central source of radiation, i.e.\ the
      variation of the normalized monochromatic flux with wavelength, 
\item the absorption and scattering efficiencies of the grains,
\item the normalized density distribution of the dust, 
\item the radius of the outer boundary in units of the inner boundary,
\item the dust temperature at the inner boundary,
\item the overall optical depth at a reference wavelength. 
\end{itemize}
For a given set of parameters, {\small DUSTY} iteratively determines
the radiation field and the dust temperature distribution by solving
an integral equation for the energy density, which is derived from a
formal integration of the radiative transfer equation. For a
prescribed radial grid the numerical integrations of radial functions
are transformed into multiplications with a matrix of weight factors 
determined purely by the geometry. Then, the energy density at every
point is determined by matrix inversion, which avoids iterations over
the energy density itself and allows a direct solution of the pure
scattering problem. Typically fewer than 30 grid points are needed to
achieve a relative error of flux conservation of less than 1\%. The
number of points used in angular integrations is 2--3 times the number
of radial grid points, and the build--in wavelength grid has 98 points
in the range from $0.01\,{\rm\mu m}$ to 3.6 cm (see Appendix C in
Ivezi{\'c} \& Elitzur \cite{IE97}). 

The distributed version of the code provides a variety of quantities
of interest including the monochromatic fluxes and the spatial
intensity distribution at wavelengths selected by the user, but not
the corresponding visibilities.
Since we want to employ the visibilities obtained from our high
spatial resolution measurements as constraints for the radiative
transfer models, we have supplemented the code with routines for the
calculation of synthetic visibility functions.

\subsection{Selection of photometric data}
\label{photometry}

An important ingredient for the radiative transfer modeling of 
circumstellar dust shells around evolved stars is the spectral energy
distribution (SED). Due to the variability of Miras and OH/IR stars,
the SED of such objects ideally has to be determined from coeval
observations covering all wavelengths of interest.
Unfortunately, no such coeval photometric data set for the wavelength
region from $\lambda \approx 1\,{\rm\mu m}$ to $\lambda \ge
20\,{\rm\mu m}$ is available in the literature for \object{AFGL
2290}. Thus, we have to define a `composite' SED, which is derived
from  observations made by different authors at different epochs, but
at about the same photometric phase (Griffin \cite{Grif93}).

\begin{table*}
\caption{Infrared photometry of \object{AFGL 2290} ordered by the date
         of observation} 
\label{IR-obstab} 
\begin{tabular}{|r|c|c|c|r|r|r|r|r|r|r|r|r|r|}\hline
No. & Julian Date  & ${\rm Phase^{*}}$ & Ref.\ &
\multicolumn{10}{c|}{Wavelengths}\\ \hline
    & 244 0000+    & $P=1424\,{\rm d}$ &       &
\multicolumn{10}{c|}{[$\,{\rm\mu m}$]}\\ \hline
1   & 1045 & 0.320 $(-3)$ & 1&
     &     &     &  4.2&     &     &     & 11.0&     & 19.8 \\
2   & 2295 & 0.198 $(-2)$ & 1&
     &     &     &     &     &     &     & 11.0&     & 19.8 \\
3   & 2725 & 0.500 $(-2)$ & 2&
     &  2.2&  3.6&  5.0&8.4, 8.8& & 10.4 10.6& 11.6& 12.6&  \\
4   & 3726 & 0.203 $(-1)$ & 3&
     &  2.3&  3.6&  4.9&  8.7&     & 10.0& 11.4& 12.6& 19.5 \\
5   & 3972 & 0.376 $(-1)$ & 4&
 1.25, 1.65&  2.2&  3.7&  4.8&  &  &     &     &      &     \\
6   & 4082 & 0.453 $(-1)$ & 3&
     &  2.3&  3.6&  4.9&  8.7&     & 10.0& 11.4& 12.6& 19.5 \\
7   & 4348 & 0.640 $(-1)$ & 3&
     &  2.3&  3.6&  4.9&  8.7&     & 10.0& 11.4& 12.6& 19.5 \\
8   & 4533 & 0.770 $(-1)$ & 4&
 1.25, 1.65&  2.2&  3.7&  4.8&  8.2&  9.6& 10.2& & 12.2&19.6\\
9   & 5146 & 0.200 $(+0)$ & 5&
     &     &  3.8&  4.8&  8.7&  9.7& 10.5& 11.5& 12.5& 20   \\
10  & 7816 & 0.075 $(+2)$ & 6&
 1.63& 2.23& 3.79&     &     &     &     &     &      &     \\
11  & 7832 & 0.087 $(+2)$ & 7&
 1.26, 1.68& 2.28& 3.80&     &     &     &     &   &  &     \\
12  & 8041 & 0.233 $(+2)$ & 8&
 1.24, 1.63& 2.19& 3.79& 4.64&     &     &     &   &  &     \\
\hline
\end{tabular}\\
References: 1) Price \& Murdock \cite{PM76},
            2) Lebofsky et al.\ \cite{LKRL76},
            3) Gehrz et al.\ \cite{GKMHG85}, 
            4) Engels \cite{Eng82}, 
            5) Herman et al.\ \cite{HISH84},
      \mbox{6) Noguchi et al.\ \cite{NQWW93}},
            7) Xiong et al.\ \cite{XChG94},
            8) Nyman et al.\ \cite{NHB93}.\\ 
${}^{*}$ Numbers in parantheses give the cycle with respect to epoch
JD 244~4860.8.
\end{table*}

From the infrared photometry of \object{AFGL 2290} available in the
literature, we selected those publications which specify the date of
observation and present the fluxes in tabulated form, either in
physical units (e.g.\ Jy) or in magnitudes with given conversion
factors (at least as a reference).
Table \ref{IR-obstab} lists the references, the date and phase of
observation and the wavelengths. The phases were determined from the
period $P = 1424$d and the epoch of maximum, JD = 244 4860.8, which
has been derived from the monitoring of the OH maser emission by
Herman \& Habing (\cite{HerHab85}). Engels et al.\ (\cite{EKSS83})
determined periods of OH/IR stars from infrared observations and
found that the periods and phases are in agreement for objects in
common with the sample Herman \& Habing (\cite{HerHab85}).

\begin{figure}\unitlength 1cm
\includegraphics{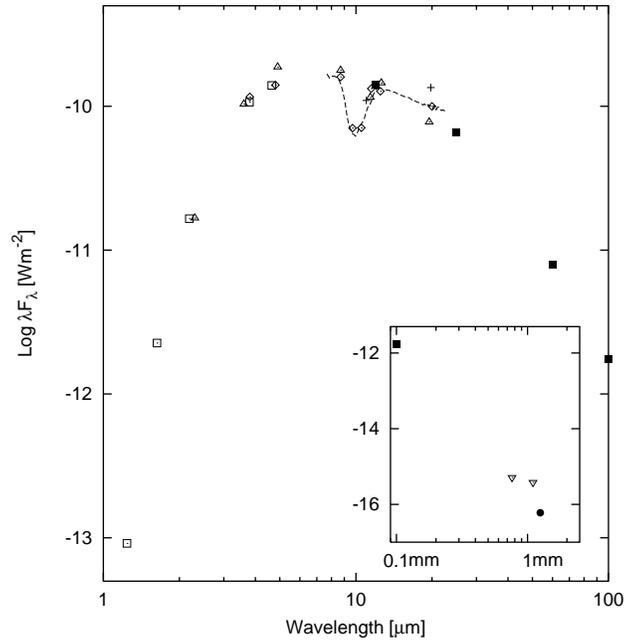}
\caption{
   IR--fluxes of \object{AFGL 2290} observed about phase 0.2 and at
   phase 0.77. The data are taken from Price \& Murdock \cite{PM76}
   ($+$), Herman et al.\ \cite{HISH84} ($\diamond $), Gehrz et al.\
   \cite{GKMHG85} ($\triangle $), and Nyman et al.\ \cite{NHB93}
   ($\sq $). 
   Also shown are the colour corrected IRAS fluxes adopted from van
   der Veen et al.\ \cite{VOHM95} (filled sqares), and the IRAS low 
   resolution spectra (dashed line). The IRAS data are multiplied by a
   factor of 1.95 in order to match the photometric data at $12\,{\rm
     \mu m}$. The insert shows mm measurements by Walmsley et al.\
   \cite{WCKSFO91} ($\bullet$) and van der Veen et al.\ \cite{VOHM95}
   ($\bigtriangledown$, $3\,\sigma$ upper limits).}
\label{IR-obsfig}
\end{figure}

It can be seen from the entries in Table \ref{IR-obstab} that
the wavelength range from  $\lambda=1.2 \,{\rm\mu m}$ to
$\lambda=20\, {\rm\mu m}$ is only fully covered by observations
around phase 0.2 (see entries 2, 4, 9, 12).The respective fluxes are
shown in Fig.\ \ref{IR-obsfig}. 

The measurements of Herman et al.\ (\cite{HISH84}) and Nyman et al.\
(\cite{NHB93}) match each other quite well at $\lambda = 3.8$ and 
$\lambda = 4.8\,{\rm\mu m}$, although the observations are separated
by two periods. The fluxes of Price \& Murdock (\cite{PM76}) and Gehrz
et al.\ (\cite{GKMHG85}) agree with the Herman et al.\ and Nyman et
al.\ data within the errors given by the authors.

To represent the SED of \object{AFGL 2290} we adopt the data of Herman 
et al.\ (\cite{HISH84}), Gehrz et al.\ (\cite{GKMHG85}), and Nyman et
al.\ (\cite{NHB93}). The scatter between the different data sets gives
a rough estimate of the uncertainty of the `composite' SED at phase
0.2 of $\approx 0.25$. We do not correct for interstellar extinction 
because the corrections are less than, or of the same order as, the
uncertainty estimated above. For \object{AFGL 2290} Herman et al.\
(\cite{HISH84}) give a value of $A_{V} = 1.6$ for an adopted distance
of 1.19 kpc which reduces to $A_{V} = 1.3$ at a distance of 0.98
kpc. With the wavelength dependence of the interstellar extinction in
the infrared from Becklin et al.\ (\cite{BNMW78}) one obtains a
correction factor of 1.35 at $\lambda = 1.25\,{\rm\mu m}$, 1.03 at
$\lambda = 4.8\,{\rm\mu m}$, and 1.15 at $\lambda = 9.5\,{\rm\mu
m}$. 

\object{AFGL 2290} was observed by IRAS (IRAS Point Source Catalog
\cite{IRAS85a}). We adopt the colour corrected broadband fluxes given
by van der Veen et al.\ (\cite{VOHM95}) and the IRAS low resolution
spectra from the IRAS Catalog of Low Resolution Spectra
(\cite{IRAS87}). The latter are corrected according to Cohen et al.\
(\cite{CWW92}). Since the broadband fluxes and spectra are averages of
several measurements taken at different phases (IRAS Explanatory
Supplement \cite{IRAS85b}), the flux levels, e.g.\ at $12\,{\rm\mu 
 m}$, are lower than the fluxes from ground based observations around
phase 0.2. Therefore, we multiply the IRAS data with a factor of 1.95
to join them with the ground based data. 

Finally, observations at mm wavelengths were reported by Walmsley et
al.\ (\cite{WCKSFO91}) who measured a flux of 0.025 Jy at $\lambda =
1.25\,{\rm mm}$ at phase 0.18 (JD 2447960), and by van der Veen et
al.\ (\cite{VOHM95}) who derived $3\sigma$ upper limits of 0.13 Jy and
0.14 Jy at 0.76 mm and 1.1 mm respectively for phase 0.25 (JD
2448069), which are consistent with the 1.25 mm  flux.

\subsection{Selection of input parameters}
\label{input-parameters}

We represent the central star by a blackbody with an effective
temperature $T_{\rm eff}$. In contrast to the visible M type Mira
variables with $T_{\rm eff} \la 3500\,{\rm K}$, the effective
temperature of OH/IR stars with optically thick dust shells cannot be
directly determined. However, if OH/IR stars can be considered as an
extension of the Mira sequence to longer periods and larger optical
depths, one might extrapolate the period--$T_{\rm eff}$ relation for
Mira variables derived by Alvarez \& Mennessier (\cite{AlMe97}) to
$P > 650\,{\rm d}$, which yields $T_{\rm eff} < 2500\,{\rm K}$ in
agreement with the values expected for the tip of the AGB. 

The dust density distribution is obtained from the velocity law, which
results from an approximate analytic solution for a stationary dust
driven wind with constant mass loss rate (e.g.\ Schutte \& Tielens
\cite{SchTi89}). If the gas pressure force is neglected and the flux
averaged absorption coefficient is assumed to be constant with the radius
$r$ in the wind, the velocity distribution is given by 
\begin{equation}\label{veleq}
 v(r) = v_{\infty} \sqrt{ 1 - \frac{r_{1}}{r}
                   \left( 1 -
                   \left(\frac{v_{1}}{v_{\infty}}\right)^{2}\right) }
\end{equation}
where $v_{1}$ denotes the velocity at the inner boundary $r_{1}$, and
$v_{\infty}$ is the velocity at infinity. The relevant free parameter
is the ratio of these velocities $\delta = v_{1}/v_{\infty}$, because
only the normalized density distribution enters into the calculation.

We adopt this velocity law, because it accounts for the changing
density gradient due to the acceleration of the matter by radiation
pressure on dust in the innermost parts of the dust shell. Compared to
a dust shell with a $1/r^{2}$ density distribution and equal optical
depth the dust density at $r_{1}$ is higher by a factor of $0.5
(1+\delta)/\delta$ and the mass loss rate is lower by a factor of $0.5
(1+\delta)$ (cf.\ Le Sidaner \& Le Bertre \cite{LSLB93}). 
According to the theory of dust driven winds the velocity at the inner
boundary, where efficient grain condensation takes place and the
acceleration of the matter by radiation pressure on dust starts, is
close to local sound velocity $c_{s}$ (see Gail \cite{Gail90}). This
is supported by observations of the velocity separation of the SiO
maser emission in OH/IR stars (Jewell et al.\ \cite{JBWW84}), which
presumably originates from the dust forming region. With $c_{s} \la
2\,{\rm kms^{-1}}$ for temperatures of  about $1000\,{\rm K}$ and with 
the measured outflow velocity of \object{AFGL 2290} of $v_{\infty} =
16\,{\rm kms^{-1}}$ (Herman \& Habing \cite{HerHab85}) one obtains
$\delta \approx 0.12$, which we adopt as the standard value for
$\delta$. 

As described in the previous section, the location of the inner
boundary $r_{1}$ of the dust shell is determined by the choice of the
dust temperature $T_{1}$ at $r_{1}$. For the outer boundary $r_{\rm
 out}$ we adopt a default value of $r_{\rm out} = 10^{3}\, r_{1}$. As
shown in the following section a larger outer boundary affects only
the far infrared fluxes for $\lambda = 100\,{\rm\mu m}$ without
altering the other properties of the model. 

We consider spherical grains of equal size described by the grain
radius $a_{\rm gr}$. This is certainly a simplification because based
on theoretical and observational arguments, one expects the presence of
a grain size distribution $n(a_{\rm gr})$. Therefore, a size
distribution similiar to the one observed in the ISM ($n(a_{\rm gr}) 
\propto a_{\rm gr}^{-3.5}$) is often assumed for radiative transfer
models of circumstellar dust shells (e.g.\ Justtanont \& Tielens
\cite{JusTi92}, Griffin \cite{Grif93}). Consistent models for {\em
stationary} dust driven winds, which include a detailed treatment of
(carbon) grain formation and growth, in fact yield a broad size
distribution which can well be approximated by a power law (Dominik et
al.\ \cite{DGS89}). However, in circumstellar shells around {\em
pulsating} AGB stars the conditions determining the condensation of
grains change periodically. The time available for the growth of the
particles is restricted by the periodic variations of the temperature
and density. This results in a narrower size distribution
(Gauger et al.\ \cite{GGS90}, Winters et al.\ \cite{WFBS97}) which
might roughly be approximated by a single dominant grain size. 

For the dust optical properties we adopt the complex refractive index
given by Ossenkopf et al.\ (\cite{OHM92}) for `warm, oxygen--deficient'
silicates. The authors consider observational determinations
of opacities of circumstellar silicates as well as laboratory data and
discuss quantitatively the effects of inclusions on the complex
refractive index, especially at shorter wavelengths ($\lambda <
8\,{\rm\mu m}$). These constants yield a good match of the overall
spectral shape of the observed SED of \object{AFGL 2290}, especially
of the $9.7\,{\rm\mu m}$ silicate feature. 
However, we will also discuss the effects on the radiative transfer
models resulting from different optical constants in the Appendix.
With the tabulated values of the complex refractive index the
extinction and scattering efficiencies are calculated from Mie theory
for spherical particles assuming isotropic scattering.\\

Once a satisfactory fit of the spectral shape is achieved with
suitably chosen values for the remaining input parameters $T_{\rm
 eff}$, $T_{1}$, $a_{\rm gr}$, and $\tau_{0.55}$ (the optical depth at
the reference wavelength $0.55\,{\rm\mu m}$) the match of the
normalized synthetic SED with the observed SED determines the
bolometric flux at earth $f_{\rm b}$. 
Combined with the effective temperature one obtains the angular
stellar diameter $\theta_{*}$ and thereby the spatial scale of the
system. With an assumed distance $D$ to the object, the luminosity
$L_{*}$, the radius of the inner boundary in cm, and the dust mass
loss rate $\dot{M}_{\rm d}$ can be calculated. The latter quantity is
given by: 
\begin{equation}\label{mdoteq}
\dot{M}_{\rm d} = 2 {\rm\pi} r_{1} v_{\infty} (1+ \delta)
                    \rho_{\rm gr}
                    \frac{\tau_{\rm d}(\lambda)}
                         {\tilde{Q}_{\rm ext}(\lambda)}
\end{equation}
where $\rho_{\rm gr}$ denotes the specific density of the grain
material, $\tau_{\rm d}(\lambda)$ is the dust optical depth at
wavelength $\lambda$, and $\tilde{Q}_{\rm ext} = \kappa_{\rm ext}
/ V_{\rm gr}$ is the extinction cross section $\kappa_{\rm ext}$ per
unit volume $V_{\rm gr}$ of the grains. For single sized grains 
$\tilde{Q}_{\rm ext}$ is proportional to the extinction efficiency
divided by the grain radius $Q_{\rm ext} / a_{\rm gr}$, and in the
Rayleigh limit $2 \pi a_{\rm gr} \ll \lambda$ it is independent of the
grain radius $a_{\rm gr}$. For the specific density of silicate grains
we adopt $\rho_{\rm gr} = 3 {\rm gcm^{-3}}$ as a typical value. For
the distance to \object{AFGL 2290} we use $D = 0.98\,{\rm kpc}$ (van
Langevelde et al.\ \cite{LHS90}).


\section{Radiative transfer modeling of \object{AFGL 2290}}
\label{radtrf-modeling}

\subsection{A radiative transfer model for the SED of \object{AFGL 2290}} 
\label{best-fit-sed}

Starting with the parameters of previous radiative transfer models for
\object{AFGL 2290} presented by Rowan--Robinson (\cite{RowRob82}),
Bedijn (\cite{Bed87}), and Suh (\cite{Suh91}), we achieved a
satisfactory match of the observed SED with the set of parameters
given in Table \ref{tab-14} after a few trials. Henceforth we will
refer to this parameter set as model A, and we will first discuss its
properties before we use it as a reference for an investigation of the
sensitivity of the results on the parameters. 

\begin{table}
\caption{Parameters and resulting properties of model A.}
\label{tab-14}
  \begin{tabular}{ccccc}\hline
   $T_{\rm eff}$ [K]  & $ T_{1}$ [K] & $a_{\rm gr}$ [${\rm\mu m}$] &
   $r_{\rm out}/r_{1}$& $\tau_{0.55}$\\
   2000 &  800 & 0.10 & $10^{4}$ & 100\\ \hline
        &      &      &          &     \\
   $\dot{M}_{\rm d}$ [${\rm M_{\odot}yr^{-1}}$]& $r_{1}$ [$R_{*}$]  &
   $f_{\rm b}$ [${\rm W m^{-2}}$] & $\theta_{*}$ [mas]& $\tau_{10}$\\
   2.7 $10^{-7}$ & 7.80 & 3.0 $10^{-10}$ & 7.50 & 7.49  \\ \hline
  \end{tabular}
\end{table}

\begin{figure}\unitlength 1cm
\includegraphics{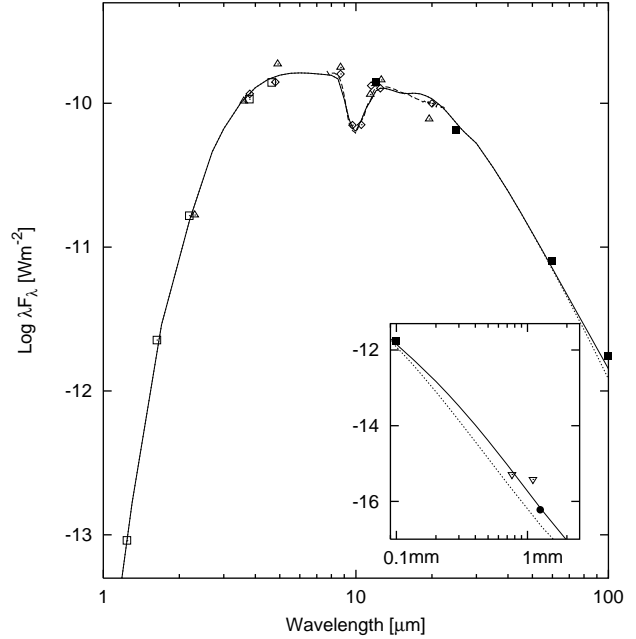}
\caption{
   Spectral energy distribution for model A (solid line) and adopted
   photometry of \object{AFGL 2290} at about phase 0.2 (see Fig.\
   \ref{IR-obsfig} for the references and corresponding symbols). The
   insert shows the SED in the mm range. The dotted line displays the
   SED for a model with the parameters of model A and an outer
   boundary of $r_{\rm out}=10^{3}\,r_{1}$.}  
\label{sed-14}
\end{figure}

\begin{figure}\unitlength 1cm
\includegraphics{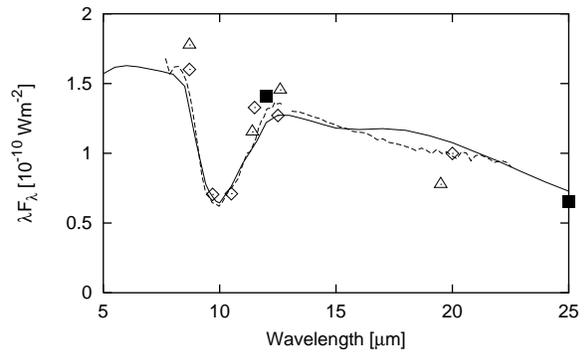}
\caption{
   $5-25\,{\rm\mu m}$ spectrum of model A (solid line) and the
   corrected IRAS LRS spectra (dashed lines). The IRAS data are scaled
   by a factor of 1.95 to match the ground based photometry (see Fig.\
   \ref{IR-obsfig} for the references and corresponding symbols).} 
\label{lrs-14}
\end{figure}

The SED of model A is shown in Fig.\ \ref{sed-14}. Figure \ref{lrs-14}
displays an enlargement of the $5-25\,{\rm\mu m}$ region with the
$9.7\,{\rm\mu m}$ silicate feature. From the shortest wavelength at
$\lambda = 1.25\,{\rm\mu m}$ up to $\lambda = 1.25\,{\rm mm}$ model A
provides a good fit to the observations. The location, shape and
strength of the silicate feature around $10\,{\rm\mu m}$ is well
reproduced with the adopted optical data from Ossenkopf et al.\
(\cite{OHM92}). Only in the $18\,{\rm\mu m}$ region is there a
noticeable deviation because the model shows a weak, broad emission
which is absent in the IRAS LRS spectrum. 

In addition to the input parameters, Table \ref{tab-14} also
lists the derived properties of model A. 
Our value for the bolometric luminosity at earth of $f_{\rm b} = 3\,
10^{-10}\,{\rm W m^{-2}}$ is consistent with the values of $2.4\,
10^{-10}\,{\rm W m^{-2}}$ determined by van der Veen \& Rugers
(\cite{VeRu89}). With $T_{\rm eff}=2000\,{\rm K}$ the angular stellar
diameter is $\theta_{*}=7.5\,{\rm mas}$ and we obtain a stellar radius 
$R_{*}=790\,{\rm R_{\odot}}$ and a luminosity $L_{*}=9000\,{\rm
 L_{\odot}}$ adopting a distance $D=0.98\,{\rm kpc}$.  
The bolometric flux should be quite accurate considering the quality
of the fit. However, the errors in the distance determination (e.g.\
$\sigma_{D}=0.34$, van Langevelde et al.\ \cite{LHS90}) and the
uncertainty in the determination of $T_{\rm eff}$ from the radiative
transfer modeling (see Sect.\ \ref{Teff-effects}) are rather large,
resulting in correspondingly large uncertainties for $L_{*}$, $R_{*}$
and $\dot{M}_{\rm d}$.

The derived dust mass loss rate of $2.7\, 10^{-7}\,{\rm
M_{\odot} yr^{-1}}$ is close to the values derived from radiative
transfer models by other authors. Bressan et al.\ (\cite{BGS98})
obtain $4.5\, 10^{-7}\,{\rm M_{\odot} yr^{-1}}$, and the model of
Bedijn (\cite{Bed87}) yields\footnote{
   We have calculated $\dot{M}_{\rm d}$ from the values for $r_{1}$
   and $\tau_{10}$ given by Bedijn (\cite{Bed87}) adopting
   $\tilde{Q}_{\rm ext}(10\,{\rm\mu m})=5.6\,10^{3}\,{\rm cm^{-1}}$.}
$4\, 10^{-7}\,{\rm M_{\odot}yr^{-1}}$.
From a relation between the strength of the $10\,{\rm\mu m}$ feature
and the color temperature Schutte \& Tielens (\cite{SchTi89}) obtained
$\dot{M}_{\rm d} = 2.4\, 10^{-7}\,{\rm M_{\odot} yr^{-1}}$ for
\object{AFGL 2290}, and Heske et al.\ (\cite{HFOVH90}) estimated
$\dot{M}_{\rm d} = 1.2\,10^{-7}\,{\rm M_{\odot} yr^{-1}}$ from the
$60\,{\rm\mu m}$ IRAS flux. 

All together the parameters and derived properties of model A lie in
the typical range of values obtained from radiative transfer models
for OH/IR stars showing the silicate feature in absorption. In the
calculations of Lorenz--Martins \& de Ara{\'u}jo (\cite{LMA97})
$T_{\rm eff}$ ranges from $1800\,{\rm K}$ to $2400\,{\rm K}$, $T_{1}$
from $650\,{\rm K}$ to $1200\,{\rm K}$, and $\tau_{9.7}$ from 7 to 17.
Justtanont \& Tielens (\cite{JusTi92}) derive dust mass loss rates
between $2.6\, 10^{-7}\,{\rm M_{\odot}yr^{-1}}$ and
$2.2\,10^{-6}\,{\rm M_{\odot}yr^{-1}}$, and optical depths
$\tau_{9.7}$ between 4.85 and 19.6 for their sample of OH/IR stars.

In addition to model A with $r_{out}/r_{1} = 10^{4}$, Fig.\
\ref{sed-14} also displays the SED for a model with the same
parameters but with an outer boundary, which is ten times smaller,
i.e.\ $r_{out}/r_{1} = 10^{3}$ (dotted line). 
The spectra are virtually indistinguishable up to $\lambda \ga 60
{\rm\mu m}$. The additional cold dust due to the larger outer
boundary of model A increases the far infrared fluxes (see insert of
Fig.\ \ref{sed-14}), but does not affect the SED at shorter
wavelengths (cf.\ Justtanont \& Tielens \cite{JusTi92}). Since the
values of the derived properties ($f_{\rm b}$, $\dot{M}_{\rm d}$, ...)
are determined essentially by the shape of the SED below $60\,{\rm\mu
m}$, they do not change for $r_{out}/r_{1} \ga 10^{3}$. 

\begin{figure}\unitlength 1cm
\includegraphics{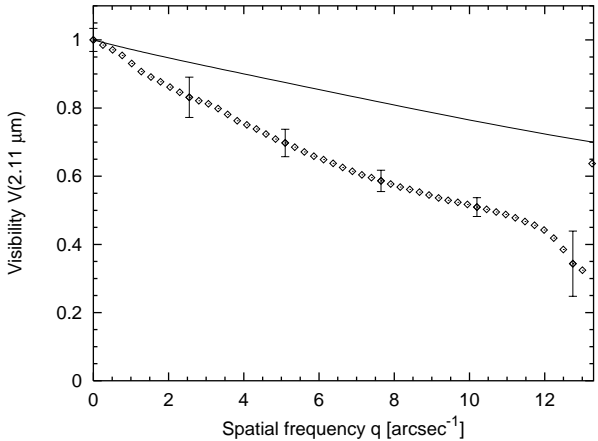}
\caption{
   Synthetic $2.11\,{\rm\mu m}$ visibility for model A (solid line)
   and the azimuthally averaged $2.11\,{\rm\mu m}$ visibility obtained
   from the speckle observation (squares).}
\label{vis-14}
\end{figure}

Figure \ref{vis-14} shows a comparison of the $2.11\,{\rm\mu m}$
visibility calculated for model A with the azimuthally averaged
$2.11\,{\rm\mu m}$ visibility $V_{\rm obs}$ derived from our speckle
observations. Although model A yields a good fit to the observed SED, 
it fails to reproduce the observed visibility. The slope of the model
visibility is much shallower than the slope of the observed one, and 
the model visibility  levels off at $q \ga 20\,{\rm arcsec^{-1}}$ with
$V_{2.11}=0.6$. From the observed visibility we obtained an upper limit
for the contribution of the star to the monochromatic flux at
$2.11\,{\rm\mu m}$ of $\approx 40\,\%$, whereas model A yields
$62\,\%$, indicating that the model optical depth of $\tau_{2.11}=3$
at $2.11\,{\rm\mu m}$ is too low. Furthermore the steeper slope of
$V_{\rm obs}$ suggests that a levelling--off should occur at a smaller
spatial frequency compared to the model visibility, which means that
the observed intensity distribution would be rather more extended
than the intensity profile of model A.

To summarize, a dust shell model, which yields a visibility in
agreement with the observations, requires a larger optical depth at
$2.11\,{\rm\mu m}$  and a more extended intensity distribution
compared to model A, but it has to produce the same spectral energy
distribution. It turns out that these requirements can be partially
fulfilled with suitably modified parameter values.

\subsection{Effects of parameter variations on the calculated SED and
            the model visibility}
\label{parameter-study}

In order to find a model which matches both the observed SED and the
visibility, as well as exploring the sensitivity of the SED and
the $2.11\,{\rm\mu m}$ visibility of the models on the input
parameters, we have calculated various sequences of models. In each
sequence one parameter was varied within a certain range, while the
other parameters were kept fixed at the values of model A, except for
the optical depth at the reference wavelength $\tau_{0.55}$. This 
quantity was adjusted for each model in order to obtain a match of the
SED with the observation, especially with the observed strength of the
$9.7\,{\rm\mu m}$ silicate feature. It turns out that models, for
which the SED fit the feature equally well have (almost) equal dust
mass loss rates, i.e.\ with our procedure we force the models
in a sequence to have equal mass loss rates instead of equal
$\tau_{0.55}$.

\subsubsection{Effects of different dust temperatures at the inner
               boundary} 
\label{T1-effects}

\begin{figure}\unitlength 1cm
\includegraphics{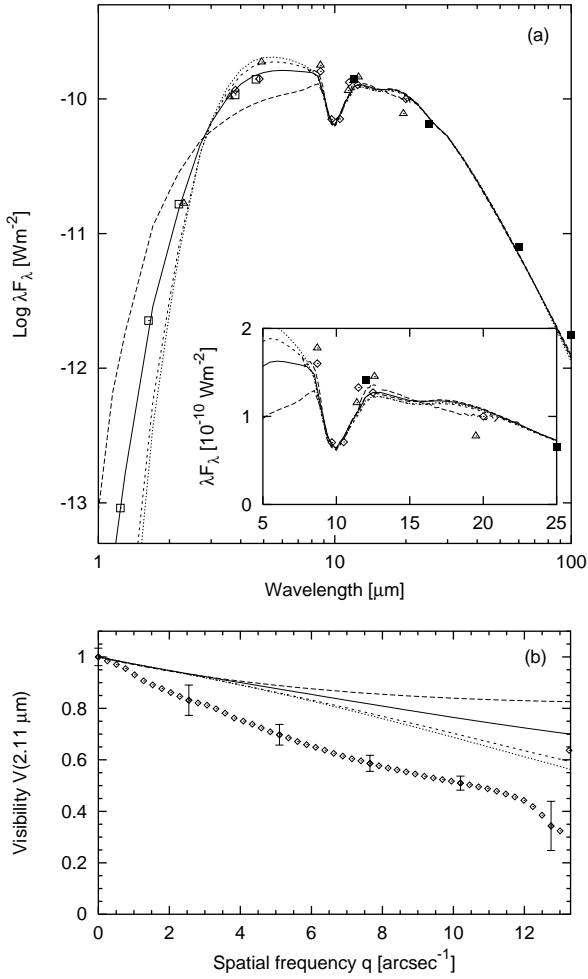}
\caption{
   SED (a) and visibilities (b) for models with the parameters of
   model A, but with different dust temperatures at the inner
   boundary:  
   $T_{1}=600\,{\rm K}$ (long dashed line),
         $800\,{\rm K}$ (solid line),
        $1000\,{\rm K}$ (short dashed line),
        $1200\,{\rm K}$ (dotted line).
   The corresponding optical depths and derived model properties are
   given in Table \ref{tab-Ow92-T1}. The spectra have been scaled with
   different $f_{\rm b}$ to match the observations at $\lambda >
   8\,{\rm\mu m}$. The symbols denote the observations (see
   Figs.\ \ref{Visibility} and \ref{IR-obsfig}).}
\label{sed-Ow92-T1}
\end{figure}

The effects of different dust temperatures at the inner boundary on
the calculated SED and the $2.11\,{\rm\mu m}$ visibility are displayed
in Fig.\ \ref{sed-Ow92-T1}. 
The dust temperature is varied between $T_{1}=600\,{\rm K}$ and
$T_{1}=1200\,{\rm K}$, and the derived properties of the corresponding
models are given in Table \ref{tab-Ow92-T1}.

\begin{table}
\caption{Resulting properties for models with the parameters of model
         A, but with different dust temperatures at the inner
         boundary.} 
\label{tab-Ow92-T1} 
 \begin{center} 
  \begin{tabular}{crcrrc}\hline
   $T_{1}$                   & $\tau_{0.55}$     &
   $\dot{M}_{\rm d}$             & $r_{1}$           &
   $\tau_{10}$               & $f_{\rm b}$            \\
   {}[K]                     &                   &
   $[{\rm M_{\odot}yr^{-1}}$]& [$R_{*}$]         &
                             & [${\rm W m^{-2}}$] \\ \hline
    600 &  65 & 2.86 $10^{-7}$ & 12.89 &  4.87 & $2.60\,10^{-10}$ \\
    800 & 100 & 2.66 $10^{-7}$ &  7.80 &  7.49 & $3.00\,10^{-10}$ \\
   1000 & 150 & 2.70 $10^{-7}$ &  5.27 & 11.23 & $3.12\,10^{-10}$ \\
   1200 & 200 & 2.66 $10^{-7}$ &  3.91 & 14.97 & $3.25\,10^{-10}$ \\
   \hline
  \end{tabular}
 \end{center} 
\end{table}

For the values of $T_{1}$ presented here, the SED is changed at
wavelengths below  $\lambda \approx 9\,{\rm\mu m}$. Compared
to model A with $T_{1}=800\,{\rm K}$ a smaller value of $T_{1}$
results in a higher flux below  $\lambda=3\,{\rm\mu m}$ and a
deficiency of flux at wavelengths between $3\,{\rm\mu m}$ and
$9\,{\rm\mu m}$. Values of $T_{1}>800\,{\rm K}$ produce a deficiency
of flux below $\lambda=3\,{\rm\mu m}$ and an excess of flux between
$3\,{\rm\mu m}$ and $9\,{\rm\mu m}$. As a consequence, the
bolometric fluxes are different for these models.

The changes of the SED for a variation of $T_{1}$ with fixed other
parameters, including $\tau_{0.55}$, have been studied already by
Ivezi{\'c} \& Elitzur (\cite{IE97}). For high $\tau_{0.55} > 100$
decreasing $T_{1}$ lowers the flux at shorter wavelengths, increases
the strength of the silicate feature and raises the flux at larger
wavelengths, similiar to the changes produced by increasing
$\tau_{0.55}$. Since we have adjusted $\tau_{0.55}$ for each value of
$T_{1}$ to reproduce the observed strength of the $9.7\,{\rm\mu m}$
silicate feature, the model SED are only changed at shorter
wavelengths in a modified way. This behaviour can be understood from
the competition of the effects caused by lowering $T_{1}$ and
simultaneously decreasing $\tau_{0.55}$. Up to $\lambda \la 3\,{\rm
\mu m}$ the increase of the monochromatic flux induced by the smaller
optical depth more than compensates the decrease of the flux due to a
lower $T_{1}$, but in the region $3\,{\rm\mu m} \la \lambda \la
8\,{\rm  \mu m}$ the effect of lowering $T_{1}$ dominates and the
monochromatic flux decreases with $T_{1}$.

The variation of $T_{1}$ mainly affects the slope of the $2.11\,{\rm
 \mu m}$ visibility $V_{2.11}$ via the different $\tau_{0.55}$. The
optical depth increases with $T_{1}$ resulting in a broader intensity
profile and a smaller stellar contribution to the monochromatic
flux. This leads to a steeper decline of $V_{2.11}$ (see Ivezi{\'c}
\& Elitzur \cite{IE96}). The curvature of the visibility only noticeably
changes at higher values of the spatial frequency. This change of
the slope of the visibility indicates the onset of a levelling off of
$V_{2.11}$, especially for the model with $T_{1}=600\,{\rm K}$. Since
its inner boundary is located at $12.9 R_{*}$ or 97~mas, the
visibility should approach a constant value at $q \approx 13\,{\rm
 arcsec^{-1}}$. An increase of $T_{1}$ above 1000~K yields only a
marginally steeper slope of the visibility, although $\tau_{0.55}$
increases and the stellar contribution to the $2.11\,{\rm\mu m}$ flux
correspondingly decreases. However, the inner boundary is shifted
simultaneously to smaller radii, resulting in similiar slopes of
$V_{2.11}$ below $q \la 13\,{\rm arcsec^{-1}}$. 

To summarize, the changes of the SED due to a variation of $T_{1}$ can
in principle be compensated by a corresponding variation of
$\tau_{0.55}$, but the presence of the silicate feature constrains
the choice of the optical depth to values which reproduce the strength
of the observed feature, i.e. to combinations of $T_{1}$ and
$\tau_{0.55}$ which yield a fixed value for $\dot{M}_{\rm d} \propto
r_{1} \tau_{10}$. Increasing $T_{1}$ yields a steeper slope of the
$2.11\,{\rm\mu m}$ visibility. Above $T_{1}=1000\,{\rm K}$, however,
these changes are small and will not result in a model which matches
the observed visibility.

\subsubsection{Effects of different effective temperatures on the SED
               and the visibility}
\label{Teff-effects}

\begin{figure}\unitlength 1cm
\includegraphics{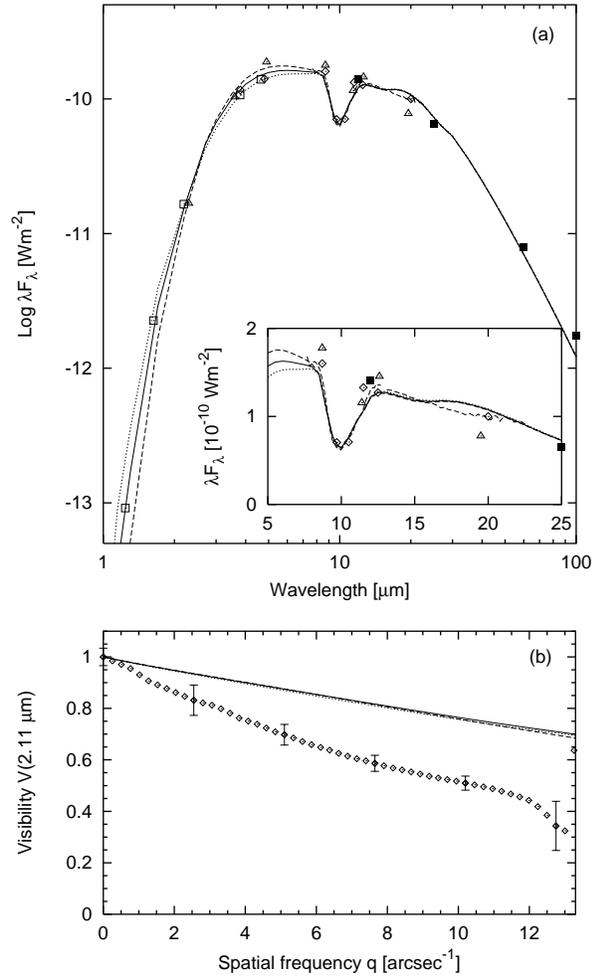}
\caption{
   SED (a) and visibilities (b) for models with the parameters of
   model A, but with different effective temperatures:
   $T_{\rm eff} = 1600\,{\rm K}$ (dashed line),
   $2000\,{\rm K}$ (solid line),
   $2400\,{\rm K}$ (dotted line).
   The corresponding optical depths and derived properties are
   given in Table \ref{tab-Ow92-Teff}. The spectra have been scaled
   with different $f_{\rm b}$ to match the observations at $\lambda >
   8\,{\rm\mu m}$. The symbols denote the observations (see
   Figs.\ \ref{Visibility} and \ref{IR-obsfig}).}  
\label{sed-Ow92-Teff}
\end{figure}

Figure \ref{sed-Ow92-Teff} displays the effects of different effective
temperatures on the calculated SED and the  $2.11\,{\rm\mu m}$
visibility. The effective temperature is varied between $T_{\rm
 eff}=1600\,{\rm K}$ and $T_{\rm eff}=2400\,{\rm  K}$. The derived
properties of the corresponding models are given in
Table \ref{tab-Ow92-Teff}.

\begin{table}
\caption{Resulting properties for models with the parameters of model
         A, but with different effective temperatures.}
\label{tab-Ow92-Teff}
 \begin{center} 
  \begin{tabular}{crcrrcc}\hline
   $T_{\rm eff}$                 & $\tau_{0.55}$     &
   $\dot{M}_{\rm d}$             & $r_{1}$           &
   $\tau_{10}$                   & $\theta_{*}$      &
   $f_{\rm b}$                      \\
   {}[K]                         &                   &
   [${\rm M_{\odot}yr^{-1}}$]    & [$R_{*}$]         &
                                 & [mas]             &
   [${\rm Wm^{-2}}$]            \\ \hline
   1600 & 110 & 2.61 $10^{-7}$ &  4.45 &  8.23 & 11.7 &
   $3.1\,10^{-10}$ \\  
   2000 & 100 & 2.66 $10^{-7}$ &  7.80 &  7.49 & 7.5  &
   $3.0\,10^{-10}$ \\
   2400 &  92 & 2.73 $10^{-7}$ & 12.60 &  6.86 & 5.2  &
   $2.9\,10^{-10}$ \\
   \hline
  \end{tabular}
 \end{center} 
\end{table}

The effects of the variation of $T_{\rm eff}$ on the model SED are
qualitatively similiar to the effects induced by a variation of
$T_{1}$. Lowering $T_{\rm eff}$ from \mbox{2400 K} to \mbox{1600 K}
decreases the monochromatic flux below $\lambda \la 2.5\,{\rm\mu m}$
and increases the flux at longer wavelengths up to $\lambda = 8\,{\rm
 \mu m}$.  The main cause for these changes is the shift of the
wavelength $\lambda_{m}$, where the stellar (blackbody) spectrum
reaches its maximum, from $\lambda_{m}=1.5\,{\rm\mu m}$ to
$\lambda_{m}=2.3\,{\rm\mu m}$. The monochromatic flux of the `hotter'
star is larger below a certain wavelength, which is, in addition,
affected by the slightly different optical depths. However, for the 
large optical depths considered here, the effects of different
$T_{\rm eff}$ on the SED are small as already shown by e.g.\
Rowan--Robinson (\cite{RowRob80}). It is interesting to note that the
changes of the SED due to a variation of $T_{1}$ can be compensated
for by a corresponding variation of $T_{\rm eff}$, at least within a
certain range of values.

The variation of $T_{\rm eff}$ has only a negligible effect on the
$2.11\,{\rm\mu m}$ visibility for the temperature range considered
here. Since the changes of the optical depths are small, both the
angular diameter of the inner boundary and the stellar contribution to
the $2.11\,{\rm\mu m}$ flux only moderately increase with
$T_{\rm eff}$. Correspondingly, the visibility approaches a slightly
higher constant value at a slightly smaller spatial frequency
resulting in the minor differences for $V_{2.11}$ at spatial
frequencies $q<13.5\,{\rm arcsec^{-1}}$. Thus, changing the
$T_{\rm eff}$ within a reasonable range cannot produce a model, which
matches the observed visibility.

\subsubsection{Effects of different grain radii} 
\label{a0-effects}

\begin{figure}\unitlength 1cm
\includegraphics{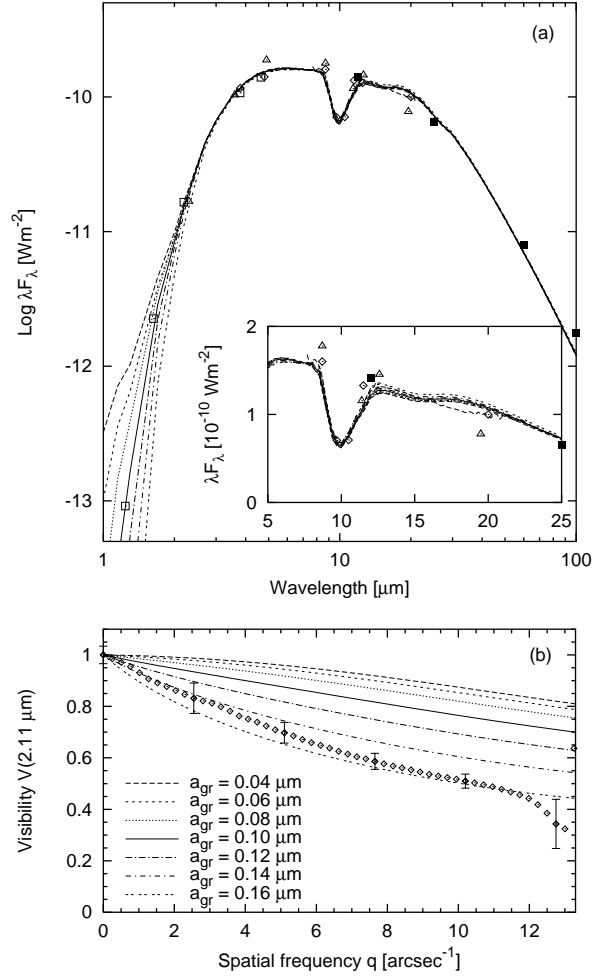}
\caption{
   SED (a) and visibilities (b) for models with the parameters of
   model A, but with different grain radii ranging from 
   $a_{\rm gr}=0.04\,{\rm\mu m}$ to $a_{\rm gr}=0.16\,{\rm\mu m}$.
   The corresponding optical depths and derived model properties are
   given in Table \ref{tab-Ow92-a0}. The symbols denote the
   observations (see Figs.\ \ref{Visibility} and \ref{IR-obsfig}).} 
\label{sed-Ow92-a0}
\end{figure}

\begin{figure}\unitlength 1cm
\includegraphics{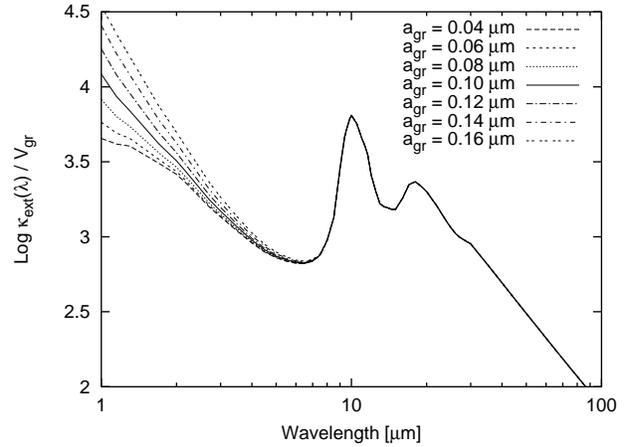}
\caption{
   Extinction coefficient per unit volume of the grains
   $\kappa_{\rm  ext}/V_{\rm gr}$ for different grain radii ranging
   from $a_{\rm gr}=0.04 \,{\rm\mu m}$ to
   $a_{\rm gr}=0.16 \,{\rm\mu m}$ calculated with the optical data 
   from Ossenkopf et al.\ (\cite{OHM92}) for `warm' silicates.}
\label{sed-Ow92-a0-qext}
\end{figure}

The effects of different grain radii on the calculated SED and the
$2.11\,{\rm\mu m}$ visibility are displayed in Fig.\ \ref{sed-Ow92-a0}.
The grain radius is varied between $a_{\rm gr}=0.04\,{\rm\mu m}$ and
$a_{\rm gr}=0.12\,{\rm\mu m}$. The derived properties of the
corresponding models are given in Table \ref{tab-Ow92-a0}. Figure
\ref{sed-Ow92-a0-qext} shows the extinction coefficient per unit 
volume of the grains $\kappa_{\rm ext}/V_{\rm gr}$ obtained from the
optical data for `warm' silicates from Ossenkopf et al.\
(\cite{OHM92}). 

The choice of the grain radius only affects the short wavelength tail
of the SED below $\lambda \la 3\,{\rm\mu m}$, because at these
wavelengths $\kappa_{\rm ext} / V_{\rm gr}$ still depends on $a_{\rm
 gr}$, but it becomes independent of $a_{\rm gr}$ at longer
wavelengths (see Fig.\ \ref{sed-Ow92-a0-qext}). This behaviour
is caused by two factors: the contribution of scattering to
extinction and the dependence of the absorption efficiency on the
grain size. The scattering efficiency per unit volume of the grains,
which is $\propto a_{\rm gr}^3$,  steeply declines with increasing
wavelength and can be neglected above a certain wavelength depending
on $a_{\rm gr}$. The absorption efficiency depends on the grain size
only at short wavelengths and becomes independent of $a_{\rm gr}$ once
the grains are sufficiently small compared to the wavelength.
Therefore, the grain radius is constrained by the observed fluxes at
the shortest wavelengths $\lambda \la 2\,{\rm\mu m}$. In our case the
photometry at $\lambda=1.65\,{\rm\mu m}$ excludes grain radii $a_{\rm
  gr} \ga  0.12\,{\rm\mu m}$ and $a_{\rm gr} \la 0.06\,{\rm\mu m}$
and the photometry at $\lambda=1.25\,{\rm\mu m}$ restricts the grain
radii to values close to $a_{\rm gr} = 0.1\,{\rm\mu m}$. However, the
values of the absorption and scattering efficiency depend on the
adopted optical data. A different data set can result in vastly
different grain radii (see Appendix A).

\begin{table}
\caption{Resulting properties for models with the parameters of model
         A, but with different grain radii.}
\label{tab-Ow92-a0}
 \begin{center} 
  \begin{tabular}{cccccc}\hline
   $a_{\rm gr}$    & $\tau_{0.55}$ & $\dot{M}_{\rm d}$ &
   $r_{1}$         & $\tau_{10}$   & $\tau_{2.2}$      \\ 
   $[ \mu $m]      &               &[${\rm M_{\odot}\,yr^{-1}}$]&
   [$R_{*}$]       &               &                   \\ \hline
   0.04 & 20.0 & 2.60 $10^{-7}$ & 6.93 & 8.22 & 2.86 \\
   0.06 & 37.5 & 2.69 $10^{-7}$ & 7.15 & 8.24 & 2.96 \\
   0.08 & 65.6 & 2.66 $10^{-7}$ & 7.41 & 7.86 & 2.98 \\
   0.10 & 100  & 2.66 $10^{-7}$ & 7.80 & 7.49 & 3.07 \\
   0.12 & 140  & 2.63 $10^{-7}$ & 8.22 & 7.03 & 3.20 \\
   0.14 & 170  & 2.60 $10^{-7}$ & 8.70 & 6.59 & 3.40 \\
   0.16 & 150  & 2.65 $10^{-7}$ & 9.24 & 6.33 & 3.76 \\ \hline
  \end{tabular}\\
  $f_{\rm b}=3\,10^{-10}\,{\rm W m^{-2}}$, $\theta_{*} = 7.5$ mas.
\end{center} 
\end{table}

The variation of the grain radius has two effects on the $2.11\,{\rm
 \mu m}$ visibility. First, the slope of $V_{2.11}$ steepens with
increasing values of $a_{\rm gr}$, because models with larger grain
radius require a higher optical depth at $2.11\,{\rm\mu m}$ in order
to match the observed SED for $\lambda>2\,{\rm\mu m}$. With
increasing optical depth the intensity distribution becomes broader
and the stellar contribution to the monochromatic flux at $2.11\,{\rm
\mu m}$ decreases. Correspondingly, the decline of visibility with
spatial frequency becomes steeper (see Ivezi{\'c} \& Elitzur
\cite{IE96}). The second effect is the change of the curvature of
$V_{2.11}$, which is noticable at low spatial frequencies.
The curvature changes its sign at about $a_{\rm gr}<0.1\,{\rm\mu m}$.
This behaviour reflects the changes of the
spatial intensity distribution. At large radial offsets $b$ from the
star the intensity decreases approximately as $I(b) \propto b^{-3}$,
because the optical depth along the line of sight at b becomes small
(see Jura \& Jacoby \cite{JuJa76}). For smaller offsets $b$, however,
the decline of the intensity steepens and the slope of the visibility
changes accordingly, i.e.\ the curvature of V indicates the changing
slope of the spatial intensity distribution. Thus, the visibility
constrains the grain radii not only via its slope, but also via its
curvature. 

The observed visibility $V_{\rm obs}$ declines in almost a straight
line to values below $\sim 0.40$ at $q=13.5{\rm arcsec^{-1}}$ with
only a slight curvature. $V_{\rm obs}$ is fairly well matched
by the model with $a_{\rm gr}=0.16\,{\rm\mu m}$, although the
curvature of the model visibility is a little too strong. Since the
visibilities for models with $a_{\rm gr}=0.15\,{\rm\mu m}$ and
$a_{\rm gr}=0.17\,{\rm\mu m}$ already fall outside the error bars of
the  observation, at least for certain spatial frequencies, the grain
radius is determined by $V_{\rm obs}$ with an uncertainty
$<0.01\,{\rm\mu m}$ (cf.\ Groenewegen \cite{Groe97}).

However, the SED for the model with $a_{\rm gr}=0.16\,{\rm\mu m}$,
shows a deficit of flux below $\lambda = 3\,{\rm\mu m}$.
This deficit cannot be removed by a change of $T_{1}$ or
$T_{\rm eff}$. Although lowering $T_{1}$ increases the flux at
$\lambda \la 3\,{\rm\mu m}$ it also decreases the flux at longer
wavelengths. Furthermore, the inner boundary of the dust shell is
moved outwards which deforms the resulting visibility in a way that
destroys the fit. Increasing $T_{\rm eff}$ yields a similiar
behaviour. 

The deficit of the short wavelength model flux could have several
causes, but the clear evidence for a non-spherical dust distribution
around \object{AFGL 2290} from our speckle masking observations favors
the explanation that the deficiency of flux in the model is due to the
assumption of a spherically symmetric circumstellar dust shell.

A more general assumption would be that the CDS has an axisymmetric,
`disk--like' structure. Theoretical investigations show that the
variation of the effective optical depth with the inclination of a
disk--like dust distribution affects the shape of the SED up to far
infrared wavelengths, as well as the monochromatic intensity
distributions and the corresponding visibilities (e.g.\ Efstathiou \&
Rowan--Robinson \cite{EfRo90}; Collison \& Fix \cite{CoFi91};
Lopez et al.\ \cite{LML95}; Men'shchikov \& Henning \cite{MeHe97}). If
the disk--like dust distribution is viewed at an intermediate
inclination one expects more flux at visual wavelengths than in the
case of a spherically symmetric dust distribution, due to scattered
light escaping from the optically thinner polar region located either
above or below the equatorial plane. In other words, we expect a
deficiency of the model flux at short wavelengths if we model the SED
of an aspherical dust distribution assuming a spherically symmetric
dust distribution.


\section{Summary and conclusions}
\label{summary}

We have presented the first diffraction--limited $2.11\,{\rm\mu m}$
speckle masking observations of the circumstellar dust shell around
the highly obscured type II OH/IR star \object{AFGL 2290}. The
resolution achieved with the SAO 6 m telescope is 75~mas, which is
sufficient to partially resolve the  circumstellar dust shell at this
wavelength. From a 2--dimensional Gaussian fit of the visibility
function the diameter was determined to be 43~mas$\times$51~mas, which
corresponds to a diameter of 42~AU$\times$50~AU for a distance of
0.98~kpc. The reconstructed image shows deviations from a spherical
structure with an elongation at position angle $130^{\circ}$. 

Our new high resolution spatial measurements provide additional strong 
constraints for radiative transfer models for the dust shell of
\object{AFGL 2290}, supplementing the information provided by the
spectral energy distribution (SED). In order to investigate the
structure and the properties of the circumstellar dust shell we have
performed radiative transfer calculations assuming a spherically
symmetric dust distribution. The spectral energy distribution at phase
$\sim 0.2$ can be well fitted at all wavelengths by a model with an
effective temperature of 2000~K, a dust temperature at the inner
boundary of 800~K, an optical depth at $0.55\,{\rm\mu m}$ of 100, and
a radius for the single--sized grains of $0.1\,{\rm\mu m}$, using the
optical constants for `warm' silicates from Ossenkopf et al.\
(\cite{OHM92}). From this fit we derived e.g.\ a bolometric flux at
earth of $3\, 10^{-10}\,{\rm W m^{-2}}$, a radius of the inner
boundary of the dust shell of $r_{1} = 7.8\, R_{*}$, and  a dust mass
loss rate of $2.7\,10^{-7}\,{\rm M_{\odot}yr^{-1}}$, in agreement with
the results of previous radiative transfer models for \object{AFGL
2290}. However, this model {\em does not} reproduce the observed
$2.11\,{\rm\mu m}$ visibility function.

We have, therefore, investigated the changes of the calculated SED and
the model visibility with the input parameters in search of an improved
model. 
We found that the grain size is the key parameter in achieving a
fit of the observed visibility, while retaining at least a partial
match of the SED. Both the slope and the curvature of the visibility 
react sensitively to the assumed grain radii.
With the assumption of single--sized grains we obtain an uncertainty
of less then $\pm 0.01\,{\rm\mu m}$ for $a_{\rm gr}$.
Another result was that the dust mass loss rate is well constrained
by the shape of the SED at longer wavelengths and, especially, by the
shape of the silicate absorption feature. For given optical constants
the value of the dust mass loss rate, as derived from the match of the
feature, is not very sensitive to changes of the input parameters.
The uncertainty of $\dot{M}_{\rm d}$ is $\sim 3\,10^{-8}\,{\rm
M_{\odot}yr^{-1}}$. The effective temperature and the dust
temperature at the inner boundary, however,  are less well
constrained. We roughly estimate a range of $\pm 300\,{\rm K}$ for
$T_{\rm eff}$ and $\pm 100\,{\rm K}$ for $T_{1}$.  

The shape of the observed visibility and the
strength of the silicate feature constrain the possible grain radii
and optical depths of the model. The observed visibility
can be reproduced by a model with a larger grain size of
$0.16\,{\rm\mu m}$ and a higher $\tau_{V}=150$, preserving
the match of the SED at longer wavelengths. Nevertheless, the model
shows a deficiency of flux at short wavelengths, which can be
explained if the dust distribution is not sperically symmetric. If the
CDS of \object{AFGL 2290} has in fact a disk--like structure, the
radial optical depths vary between the equatorial and polar
direction. Due to the scattered radiation escaping from the optically
thinner polar regions one expects more flux at shorter wavelengths
than from a spherically symmetric system with equal optical depth
towards the star.


\begin{acknowledgements}
This research made use of the SIMBAD database, operated by CDS in
Strasbourg.
\end{acknowledgements}


\appendix
\section{Effects of different optical constants}
\label{qext-effects}

In Sect.\ \ref{a0-effects} the deficiency of short wavelength flux for
the model which matches the observed $2.11\,{\rm\mu m}$ visibility of
\object{AFGL 2290} was explained by the assumption of a spherically
symmetric dust distribution. However, the deficiency could be due to a
different cause. For example, the optical constants of `astronomical'
silicates at wavelengths $\la 7.5\,{\rm\mu m}$ are not well known
(Ossenkopf et al.\ \cite{OHM92}) and it is possible that the optical
properties of the grains around \object{AFGL 2290} at short
wavelengths differ from our assumption. 

\begin{figure}\unitlength 1cm
\includegraphics{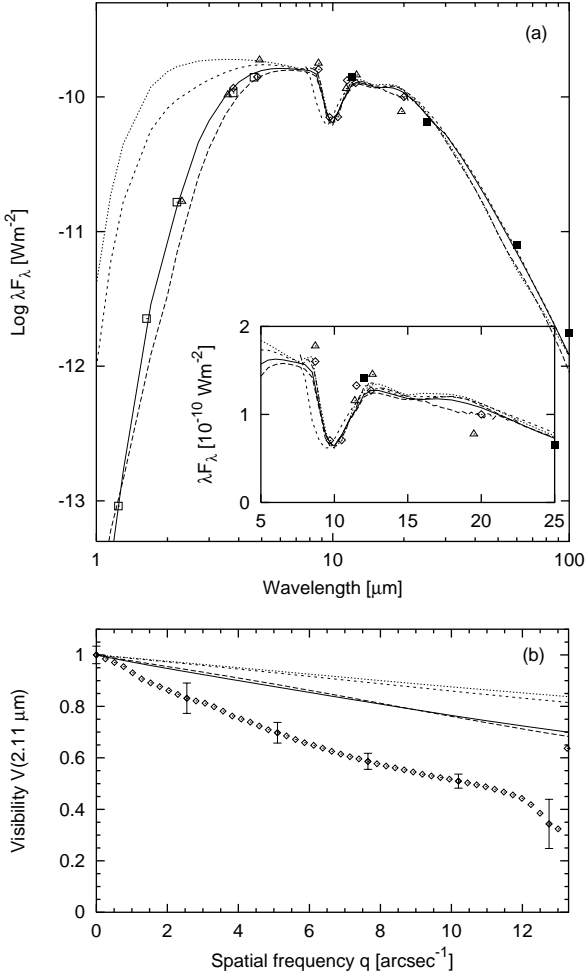}
\caption{
   SED (a) and visibilities (b) for models with the parameters of
   model A, but with different dust optical properties: 
   `warm' silicates from Ossenkopf et al.\ (\cite{OHM92}) (solid line),
   `cold' silicates from Ossenkopf et al.\ (\cite{OHM92}) (long dashed
   line), 
   Draine \& Lee (\cite{DL84}) (short dashed line), and
   David \& P{\'e}gouri{\'e} (\cite{DP95}) (dotted line).
   The corresponding optical depths and derived model properties are
   given in Table \ref{tab-Ow92-qext}. The spectra have been scaled
   with different $f_{\rm b}$ to match the observations at $\lambda >
   8\,{\rm\mu m}$.}   
\label{sed-Ow92-qext}
\end{figure}

\begin{figure}\unitlength 1cm
\includegraphics{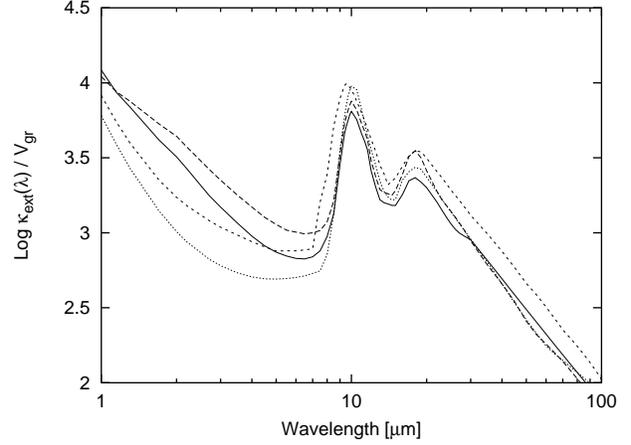}
\caption{
   Extinction coefficient per unit volume of the grains $\kappa_{\rm
   ext} / V_{\rm gr}$ for $a_{\rm gr}=0.1 \,{\rm\mu m}$ obtained from
   different optical constants: 
   `warm' silicates from Ossenkopf et al.\ (\cite{OHM92}) (solid line),
   `cold' silicates from Ossenkopf et al.\ (\cite{OHM92}) (long dashed
   line), 
   Draine \& Lee (\cite{DL84}) (short dashed line),
   David \& P{\'e}gouri{\'e} (\cite{DP95}) (dotted line).} 
\label{kap-Ow92-qext}
\end{figure}

Therefore, we have investigated the effects of different dust optical
properties on the SED and the visibility, as shown in Fig.\
\ref{sed-Ow92-qext}. The models have been calculated with the
parameters of model A (except for the value of $\tau_{0.55}$) using
the optical data from Ossenkopf et al.\ (\cite{OHM92}) for `cold'
silicates (sil--Oc), from Draine \& Lee (\cite{DL84}) (sil--DL) and
from David \& P{\'e}gouri{\'e} (\cite{DP95}) (sil--DP). The extinction
coefficient per unit volume of the grains for $a_{\rm gr} = 0.1\,{\rm
\mu m}$ is shown in Fig.\ \ref{kap-Ow92-qext} and the derived
properties of the models are given in Table \ref{tab-Ow92-qext}. 

The differences of the $\kappa_{\rm ext}/V_{\rm gr}$ resulting from
the optical data sets are more or less directly translated into
modifications of the SED, if the different values for $\tau_{0.55}$
are taken into account. Compared to the `warm' silicates of Ossenkopf
et al.\ (\cite{OHM92}) (sil--Ow), the extinction of sil--Oc grains is
higher between $\lambda \approx 1.3\,{\rm\mu m}$ and $8\,{\rm\mu
m}$, resulting in a lower monochromatic flux of the corresponding
model. For the sil--DL and sil--DP data the extinction is lower
resulting in an excess of flux. Because the shape of the silicate
features at around  $10\,{\rm\mu m}$ and  $18\,{\rm\mu m}$ is
similiar for the sil--Ow, sil--Oc, and sil--DP data, except for a
slightly different ratio of the peak strengths, they yield comparably
good fits to width and strength of the observed feature, which has its
center at $10\,{\rm\mu m}$. In contrast, the silicate feature from
the sil--DL data peaks at $9.7\,{\rm\mu m}$, and it is broader than
the observed one. 

\begin{table}
\caption{Resulting properties for models with the parameters of model
         A, but with different dust optical properties.}
\label{tab-Ow92-qext}
 \begin{center} 
  \begin{tabular}{crcccc}\hline
   Optical                   & $\tau_{0.55}$     &
   $\dot{M}_{\rm d}$         & $r_{1}$           &
   $\tau_{10}$               & $f_{\rm b}$            \\
   constants                 &                   &
   [${\rm M_{\odot}yr^{-1}}$]& [$R_{*}$]         &
                             & [${\rm W m^{-2}}$] \\ \hline
    Sil--Ow & 100 & 2.66 $10^{-7}$ &  7.80 &  7.49 & $3.00\,10^{-10}$\\
    Sil--Oc &  85 & 2.33 $10^{-7}$ &  7.63 &  7.81 & $2.75\,10^{-10}$\\
    Sil--DL &  50 & 1.57 $10^{-7}$ &  6.23 &  7.74 & $3.75\,10^{-10}$\\
    Sil--DP &  50 & 1.71 $10^{-7}$ &  5.94 &  9.29 & $4.50\,10^{-10}$\\
    \hline
  \end{tabular}
 \end{center} 
\end{table}

Because the value of $\dot{M}_{\rm d}$ depends on the adopted optical
properties of the grains (see Eq.\ \ref{mdoteq}), we obtain
$\dot{M}_{\rm d}$ for the different models ranging from
$1.6\,10^{-7}\,{\rm M_{\odot}yr^{-1}}$ to
$2.7\,10^{-7}\,{\rm M_{\odot}yr^{-1}}$ (see Table \ref{tab-Ow92-qext}).
Nevertheless, if the dust mass loss rate is derived from the match of
the silicate absorption feature, its value is not very sensitive to
variations of the effective temperature, dust temperature at the
inner boundary and the grain radius as long as the models are
calculated with the same optical constants.

The changes of the $2.11\,{\rm\mu m}$ visibilities are again caused
by the different optical depths of the models at this wavelength.
For other fixed parameters a higher optical depth produces a more
extended brightness distribution and, thereby, a steeper decline of
the visibility. The optical depth at $2.11\,{\rm\mu m}$ has similiar
values for the sil--Ow and sil--Oc models and lower, but again
similiar values for the sil--DL and sil--DP models. Hence, the decline
of visibilities from the latter models is shallower.\\

\begin{figure}\unitlength 1cm
\includegraphics{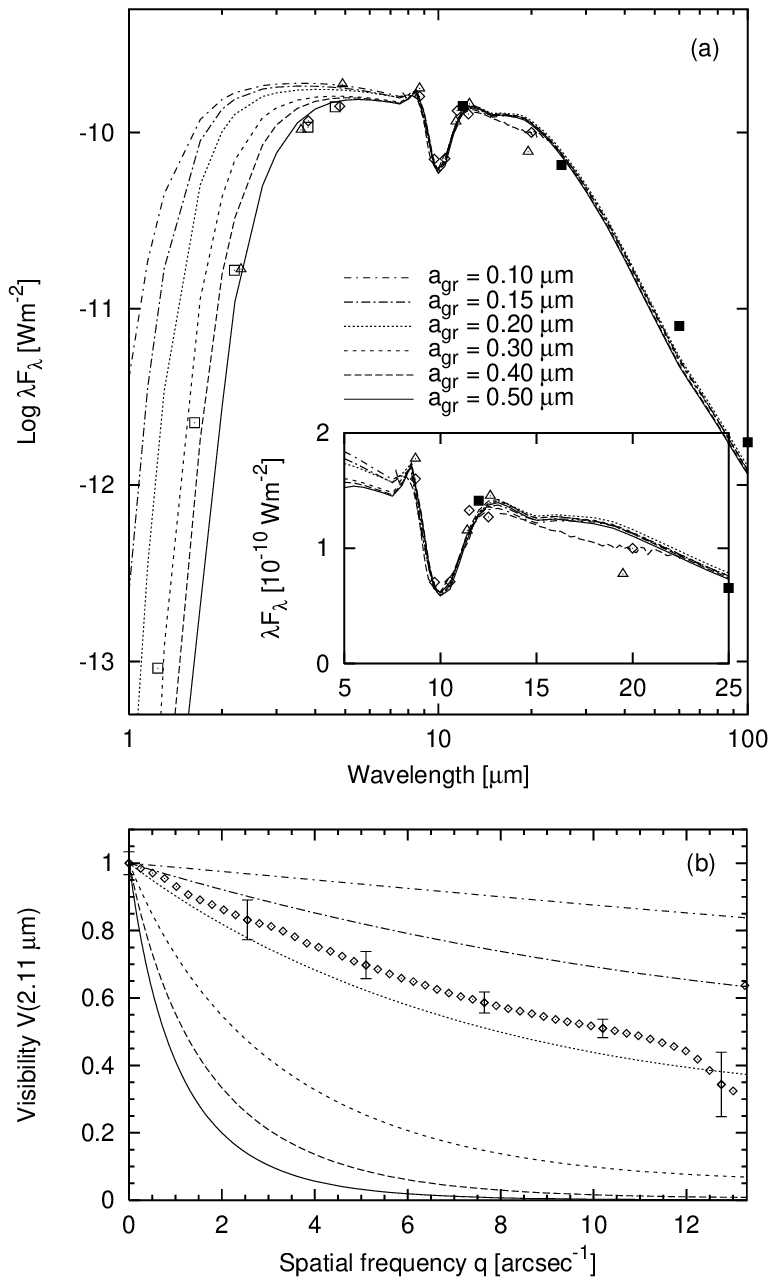}
\caption{
   SED (a) and visibilities (b) for models with
   the optical constants from David \& P{\'e}gouri{\'e} (\cite{DP95})
   and different grain radii ranging from $a_{\rm gr}=0.1 \,{\rm\mu m}$
   to $a_{\rm gr}=0.5 \,{\rm\mu m}$. The other parameters are equal to
   model A. The corresponding optical depths and derived model
   properties are given in Table \ref{tab-DP95-a0}. The spectra have
   been scaled with different $f_{\rm b}$ to match the observations at
   $\lambda >8\,{\rm\mu m}$.}
\label{sed-DP95-a0}
\end{figure}

\begin{figure}\unitlength 1cm
\includegraphics{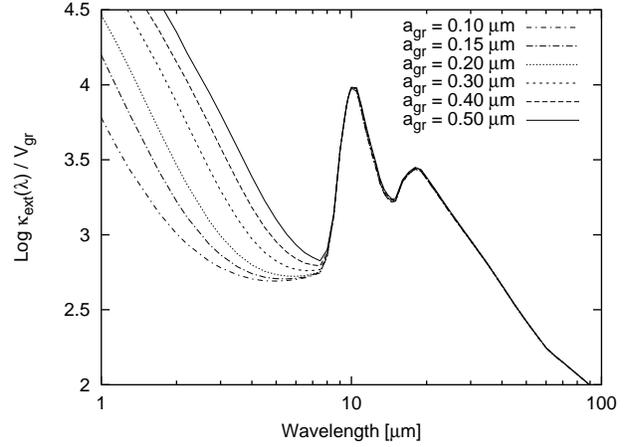}
\caption{
   Extinction coefficient per unit volume of the grains
   $\kappa_{\rm ext} / V_{\rm gr}$ calculated with the optical
   constants from David \& P{\'e}gouri{\'e} (\cite{DP95}) for
   different grain radii ranging from $a_{\rm gr}=0.1\,{\rm\mu m}$ to
   $a_{\rm gr}=0.5 \,{\rm\mu m}$.}
\label{sed-DP95-a0-qext}
\end{figure}

The optical properties from David \& P{\'e}gouri{\'e} (\cite{DP95})
yield a fit of the silicate feature, which is comparable to the fit
obtained with the Ossenkopf et al.\ (\cite{OHM92}) data, but they
produce an excess of flux at smaller wavelengths for a grain radius of
$0.1\,{\rm\mu m}$. From the investigation of the effects resulting
from a variation  of $a_{\rm gr}$ in Sect.\ \ref{a0-effects}, we know
that the flux at short wavelengths decreases with increasing grain
radius, and that the decline of the visibility becomes steeper. 
Therefore, we have calculated a series of models with the David \&
P{\'e}gouri{\'e} (\cite{DP95}) data where we varied the grain radius
from $a_{\rm gr}=0.1 \,{\rm\mu m}$ to $0.5\,{\rm\mu m}$. 
Figure \ref{sed-DP95-a0} shows the calculated SED and the 
$2.11\,{\rm\mu m}$ visibilities. The derived properties of the
corresponding models are given in Table \ref{tab-DP95-a0}.
Figure \ref{sed-DP95-a0-qext} shows the extinction coefficient per
unit volume of the grains $\kappa_{\rm ext} / V_{\rm gr}$ for the 
different grain radii.

Compared to the models calculated with the sil--Ow constants, the
sil--DP models show a flat flux distribution below $\lambda \la
8\,{\rm\mu m}$ with a steep drop at a certain wavelength, which
depends on the grain radius. Furthermore, much higher values of
$a_{\rm gr}$ are needed to match the observation. The short
wavelength photometry of \object{AFGL 2290} constrains the grain
radius to $a_{\rm gr}\sim 0.5\,{\rm\mu m}$, although the slope of the
observed SED at wavelengths $\la 2.2\,{\rm\mu m}$ is less well
reproduced compared to the sil--Ow model with $a_{\rm gr}=0.1\,{\rm\mu
m}$. The different behaviour below $\lambda =8\,{\rm\mu m}$ is caused
by the fact that the sil--DP constants yield smaller extinction
efficiencies  for grains of equal radius compared to the sil--Ow
constants. Therefore, substantially larger grains are needed to
produce comparable optical depths at shorter wavelengths. As a
consequence, the contribution of scattering to the extinction is still 
important up to wavelengths, which are approximately 3--4 times
larger.  

\begin{table}
\caption{Derived properties for models with the optical constants from
         David \& P{\'e}gouri{\'e} (\cite{DP95}) and different grain
         radii. The other parameters are equal to model A, except for
         $\tau_{0.55}$ which has been adjusted to fit the observed SED.} 
\label{tab-DP95-a0}
 \begin{center} 
  \begin{tabular}{crcrcrc}\hline
   $a_{\rm gr}$               & $\tau_{0.55}$     &
   $\dot{M}_{\rm d}$          & $r_{1}$           &
   $\tau_{10}$                & $\tau_{2.2}$      &
   $f_{\rm b}$                \\
   ${\rm [\mu m]}$            &                   &
   [${\rm M_{\odot}yr^{-1}}$] & [$R_{*}$]         &
                              &                   &
   [${\rm W m^{-2}}$] \\ \hline
   0.10 &  50 & 1.71 $10^{-7}$ &  5.94 & 9.29 & 0.85 & $3.00\,10^{-10}$ \\
   0.15 & 100 & 1.76 $10^{-7}$ &  7.00 & 8.08 & 1.12 & $3.25\,10^{-10}$ \\
   0.20 & 100 & 1.89 $10^{-7}$ &  8.10 & 7.52 & 1.72 & $3.50\,10^{-10}$ \\
   0.30 &  70 & 1.94 $10^{-7}$ &  9.71 & 6.49 & 3.74 & $4.00\,10^{-10}$ \\
   0.40 &  34 & 1.92 $10^{-7}$ & 10.52 & 5.96 & 6.61 & $4.25\,10^{-10}$ \\
   0.50 &18.6 & 2.00 $10^{-7}$ & 11.08 & 5.91 & 10.0 & $4.50\,10^{-10}$ \\
   \hline
  \end{tabular}
 \end{center} 
\end{table}

Due to the stronger dependence of $\kappa_{\rm ext} / V_{\rm gr}$ on
the grain radius (Fig.\ \ref{sed-DP95-a0-qext}) and the resulting
larger variation of the $2.11\,{\rm\mu m}$ optical depth (see Table
\ref{tab-DP95-a0}), the changes of the $2.11\,{\rm\mu m}$ visibility
are much more pronounced. With increasing $a_{\rm gr}$ the curvature
of the visibility increases, and $V_{2.11}$ falls
off to smaller values at  $q=13.5\,{\rm arcsec^{-1}}$ because the
stellar contribution to the monochromatic flux is reduced. 
Up to $a_{\rm gr}\approx 0.2\,{\rm\mu m}$ the change of the curvature
is mainly caused by the changing  slope of the spatial intensity
distribution, as discussed in Sect.\ \ref{a0-effects}. At larger grain
radii the increase of the radius of the inner boundary and the
corresponding reduction of the spatial frequencies, where $V_{2.11}$
approaches a constant value (or zero) becomes more important.

The best match of the observed visibility $V_{\rm obs}$ is obtained
for the model with $a_{\rm gr}=0.2\,{\rm\mu m}$ (dotted line in
Fig.\ \ref{sed-DP95-a0}), which is somewhat larger than $a_{\rm
 gr}=0.16\,{\rm\mu m}$ for the best matching model with the sil--Ow
optical constants. The resulting dust mass loss rate is about 
30~\% smaller, and the radius of the inner boundary is about 12~\%
smaller. Again the curvature of the model visibility is slightly too
strong. But in contrast to the model with the sil--Ow
optical constants, there is now an excess of flux at smaller
wavelengths $\lambda\la 5\,{\rm\mu m}$.

As discussed at the end of Sect.\ \ref{a0-effects}, we would expect a
deficiency of flux if the CDS of \object{AFGL 2290} has a disk--like
structure. This suggests that the optical constants from David \&
P{\'e}gouri{\'e} (\cite{DP95}) underestimate the extinction of the
grain material at short wavelengths. However, we also do not know,
whether the optical constants from Ossenkopf et al.\ (\cite{OHM92})
represent the intrinsic optical properties of the grains, because the
shape of the SED at short wavelengths is affected to an unknown degree 
by the geometry of the non-spherical dust distribution around
\object{AFGL 2290}.
For the same reason we refrain from enforcing a match of the SED at
short wavelengths by a suitable modification of the dust optical
constants in this region. 

\end{document}